\documentclass[showpacs,preprintnumbers,amsmath,amssymb]{revtex4}

\usepackage{graphicx}
\usepackage{dcolumn}
\usepackage{bm}

\def\r0{\rho_{0}}

\def\dist{{\rm dist}}

\def\a0{\alpha_0}
\def\a{\alpha}

\def\be{\begin{equation}}
\def\ee{\end{equation}}
\def\beq{\begin{equation}}
\def\eeq{\end{equation}}

\def\({\left(}
\def\){\right)}

\newcommand{\calD}{\mathcal{D}}

\newtheorem{theorem}{Theorem}

\begin{document}


\title{Nonclassical rotational inertia for a supersolid under rotation}

\author{Amandine Aftalion}\email{amandine.aftalion@math.jussieu.fr}
\affiliation{Universit\'e
Pierre et Marie Curie-Paris 6, CNRS-UMR 7598,
 Laboratoire Jacques-Louis Lions, 175 rue du Chevaleret,
  Paris, F-75013, France.}
\author{Xavier Blanc}\email{blanc@ann.jussieu.fr}
\affiliation{Universit\'e
Pierre et Marie Curie-Paris 6, CNRS-UMR 7598,
 Laboratoire Jacques-Louis Lions, 175 rue du Chevaleret,
  Paris, F-75013, France.}
\author{Robert L.Jerrard}\email{rjerrard@math.toronto.edu}
\affiliation{Department of Mathematics, University of Toronto,
Toronto, Ontario M5S 2E4, Canada.}

\date{\today}

\begin{abstract}
As proposed by Leggett \cite{L}, the supersolidity of a crystal is
characterized by the Non Classical Rotational Inertia (NCRI)
property. Using a model of quantum crystal introduced by Josserand,
Pomeau and Rica \cite{PR}, we prove that NCRI occurs. This is done
by analyzing the ground state of the aforementioned model, which is
related to a sphere packing problem, and then deriving a theoretical
formula for the inertia momentum. We infer a lower estimate for the
NCRI fraction, which is a landmark of supersolidity.
\end{abstract}

\maketitle

A recent experiment by Kim and Chan \cite{KC} allowed them to
measure the moment of inertia of solid helium and find that it is
 lower than its classical value.
 This property is referred to as Non Classical Rotational Inertia (NCRI).
  This experiment has raised a lot of interest
 and was interpreted as a landmark of supersolidity, on the basis
  of a paper by Leggett \cite{L}. In \cite{L}, Leggett
  predicted that the property of nonclassical rotational inertia possessed by
superfluid helium is shared by solids and proposed
  as a definition for the non classical rotational inertia fraction
  $\hbox{NCRIF} =(I_0-I)/I_0$ where
   $I$ is the moment of inertia of the crystal under study and $I_0$ its classical value.
    One
  theoretical challenge (see the review paper of Prokof'ev
 \cite{P}) is to estimate this NCRIF and
 check that it is nonzero. This is the aim of this paper, based on a
 model of quantum crystal, introduced by Josserand, Pomeau, Rica \cite{PR}.
 In this respect,  we derive  a key
 estimate providing the lower bound \eqref{lbn2} for the NCRIF.
 In the literature (see \cite{P}), different microscopic mechanisms have been proposed to
describe the supersolidity of a crystal, based mainly on the off diagonal long range
order property (ODLRO) of the density matrix, and Jastrow wave
functions. Here, we do not relate directly the NCRI to ODLRO, or the
presence of vacancies but choose another approach to model the
solid.

Josserand, Pomeau, Rica \cite{PR} proposed a model  of  quantum
  solid: it
 is based on
the fact that the complex valued
 wave function common to all particles of mass $m$ minimizes
 the Gross-Pitaevskii
energy with an integral term that can be viewed as a 2-body
potential in a first Born approximation:
$$\int \frac{\hbar^{2}}{2m}|{\nabla}\psi({\bm{r}})|^{2}\
d{\bm{r}}+ \frac 1 4 \int\!\!\!\int  \tilde U({\bm{r}}' - {\bm{r}})
|\psi({\bm{r}'})|^2|\psi({\bm{r}})|^2\ d{\bm{r}}\ d{\bm{r}'}
$$ where $\tilde U(\cdot)$ is a
 potential depending on the distance between atoms.
  The normalization condition is $\nu= \int
|\psi|^2/V$
 where $V$ is the volume of the region $\calD$ occupied by the solid.
 This model bears an important difference with classical solids, in
  the sense that in classical solids, there is an integer number
   of atoms per unit cell, while in this quantum solid model,
    the average density is a free number, independent of the crystal
parameters. Moreover, this model yields a dispersion relation
between the energy
  and momentum of elementary excitations that depends on the
   two-body potential. The choice of $\tilde U$ is made in order to
    have a roton minimum in this dispersion relation. For instance,
  one possibility is to take
 $\tilde U(|\bm{r}|) = U_0 \theta(a-|\bm{r}|)$, with  $\theta(.)$ the
Heaviside function \cite{PR}.
 We define $g= U_0\frac{m a^2}{\hbar^2} \nu a^3$ and rescale
 distances by $a$  so that the rescaled energy
  $E_g$ is given by\beq\label{GP}\int \frac12|{\nabla}
 \psi({\bm{r}})|^{2}\ d{\bm{r}}+ \frac g 4 \int\!\!\!\int  U({\bm{r}}' - {\bm{r}})
|\psi({\bm{r}'})|^2|\psi({\bm{r}})|^2\ d{\bm{r}}\ d{\bm{r}'}\eeq
where $U(|\bm{r}|) =  \theta(1-|\bm{r}|)$ and $ \int |\psi|^2=V$.
For small $g$, the ground state is $|\psi|=1$, and for large $g$,
computations in \cite{PR} indicate the presence of a crystal
phase with
 some supersolid-like behaviour
 under rotation. Moreover, the authors of \cite{PR} checked that this model
  also provides mechanical behaviours typical of solids
   under small stress. We believe that the model proposed in \cite{PR}
is not far from  a realistic model of solid helium, that is
 of a dense solid with
strongly repulsive interaction. Note that for He, we have $g\sim
25$, and for Ne, $g\sim 100$ \cite{pm}. In the large $g$ limit, we
will see that the ground state of (\ref{GP})
 is a periodic array of peaks. The
self interaction of a peak becomes a constant, added to the energy,
and independent of the wave function local profile. One could argue
that in a mean field model of a real crystal, the interaction has a
hard core, so that the self interaction is infinite. But in the true
physical system of solid helium,
 a given atom does not interact with itself and thus
does not provide any infinite self interaction. This, added to the
various properties of the quantum crystal derived in \cite{PR},
which are in agreement with experimental solid helium, make us
believe that the model provides insight into the understanding of
supersolids.
  The aim of this paper is to
 use this model to derive an approximate theoretical value for the
 reduction of the moment of inertia of a supersolid.
 The proof is two-fold: on the one hand, we use the specific choice
of the interaction potential $U$ to get that for large $g$ , the
ground state $\psi_g$ has a periodic density $\rho_g=\psi_g^2$.
Moreover, the wave function is localized around sets defined by a
sphere packing problem. On the other hand, given this periodic
density $\rho_g$, we use the expression (\ref{ncrif}) of the NCRIF
and the fact that $\psi_g$ is a ground state, hence a solution of
some nonlinear Schr\"odinger equation, to obtain a lower estimate of
the NCRIF (\ref{lbn}) and (\ref{lbn2}). Since this second part of
the proof only relies on the periodicity of $\rho_g$ and the fact
that it is a solution of an equation, it could be used for other
models providing a periodic density, for instance that of an
 optical lattice (with an exterior field creating a periodic density
 with several atoms per site).

\begin{figure}[bt]
\begin{centering}
\includegraphics[width=45mm]{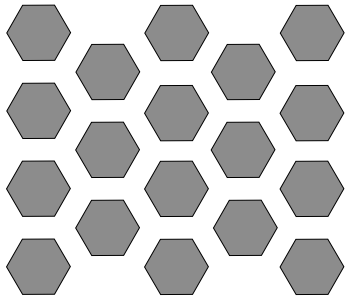}
\end{centering}
 \caption{The expected configuration of sets $A_i$ in 2D.}
 \label{fig:2}\vspace{-.5cm}
\end{figure}

 If $\calD$ is a solid sample, the sphere
packing problem \cite{CS,H} provides a number $n(\calD)$, the largest
number of points in ${\cal D}$ which are at distance larger than one
from each other.
When this
number
 is large, the optimal location
 of points is proved \cite{CS,H} to be close to a hexagonal lattice
 in 2D. In 3D, 2 configurations are optimal: body centered cubic close
 packing and face centered cubic close packing. 
 When $g$ is large, the two terms in (\ref{GP}) are
 of different  order, hence the ground state $\psi_g$ is very close to a function
 $\psi_0$ that is found by minimizing the kinetic energy within the functions that minimize the interaction term, which is dominant.  We are going to prove that
 such a function is  supported in sets $A_i$ which are
at distance at least one and whose number is
$n(\calD)$.
  Thus, the sets $A_i$ are determined by the minimization problem
  \beq\label{promin}\inf_{A_i,\ \dist( A_i, A_j)>1}\left\{ \sum_{i=1}^{n(\calD)}
 \lambda_1(A_i)\right\}\eeq
 where $\lambda_1$ is the ground energy of
 $-\Delta$ in $A_i$ with zero boundary conditions:
 $\lambda_1(A_i) = \inf_{\int |u|^2 = 1}\{\int_{A_i} |\nabla u|^2
 \}$.
 The expected configuration is illustrated in figure \ref{fig:2}.
 The function $\psi_0$ corresponds to the ground state of $-\Delta$ in each
 $A_i$ and vanishes outside the $A_i$'s. A ground state of $E_g$ will be
 very close to $\psi_0$ in the sets $A_i$, and exponentially small
 away from the $A_i$'s, except on a boundary layer.

 When the sample is set under rotation $\Omega$ about the
$z$ axis, the free energy of the system is defined
as\beq\label{eomega}e(\Omega)=\inf_{\psi} \left\{ E_g(\psi)-\Omega
\langle \psi, L_z(\psi)\rangle\right\}\eeq where $L_z(\psi)=i {\bf
r}\times \nabla\psi\cdot e_z$ and $E_g$ is the energy defined in
(\ref{GP}). When $\Omega$ is small, $e(\Omega)$ can be expanded as
$e_0-(1/2)I\Omega^2$ where $I$ is the effective moment of inertia of
the system. Leggett \cite{L} suggested as a criterion for
superfluidity the existence of a non classical rotational inertia
fraction (NCRIF), defined as $(I_0-I)/I_0$, where $I_0$ is the
classical
 moment of inertia  of the crystal phase and is equal to $\int
 |\psi_g|^2 r^2$ where $\psi_g$ is a ground state of $E_g$.
    The point of this analysis is to find an estimate for
the NCRIF,
  computed numerically in \cite{PR}, and prove that
  it is non zero. As can be seen in (\ref{ncrif}), a good knowledge
  of $\rho_g$, the ground state density with no rotation, is needed to
  estimate NCRIF in (\ref{lbn}) and
  (\ref{lbn2}).

   The paper is organized as follows: first, we study the ground
   state of the crystal phase with no rotation and derive
   (\ref{promin}). Then, we present some more refined computations
   in the 1D case, and finally we derive estimates for the NCRIF.


\noindent {\bf Crystal phase with no rotation} We first describe
the minimization  of the second term
 of (\ref{GP}) which provides a class of functions $\psi$
  such that $\rho=|\psi|^2$  has mass located in disjoint
  sets $A_i$, at distance at least the range of the potential,
  which is 1.
  When one wants to
 minimize $\int  |\nabla \psi|^2$ in this class, this provides a
 constraint (\ref{promin}) on the shape of the sets $A_i$ that we
 explain.

We denote by $(U * \rho)({\bf r})=\int   U({\bm{r}} - {\bm{r}'})
\rho({\bm{r}'}) d{\bm{r}'}$ and
  $F(\rho)=\int\int  U({\bm{r}}' - {\bm{r}})
\rho({\bm{r}'})\rho({\bm{r}})\ d{\bm{r}}\ d{\bm{r}'}$. Recall that
$n(\calD)$ was defined in the introduction and is related to the
sphere packing problem.
 \begin{theorem}\label{theo1}
 A measure $\rho$ with $\int \rho=V$
minimizes $F(\rho)$ if and only if there exist $n(\calD)$ pairwise
disjoint sets $A_1,\ldots, A_{n(\calD)}$, such that \beq
\dist(A_i,A_j) \ge 1\mbox{ if }i\ne j,\hbox{ and } \int_{A_i} \rho =
\frac {V}{n(\calD)}. \label{Mstar}\eeq Moreover, $\min F = V^2/
n(\calD)$.\end{theorem} The proof of this result which strongly
relies on the shape of $U$ is made  in the appendix. Let us call
$\rho_0$ a ground state of $F$ and $\psi_g$ a ground state of $E_g$
with $\rho_g=\psi_g^2$. Then $F(\rho_0)\leq F(\rho_g) \leq
F(\rho_0)+(1/g) \int |\nabla \psi_0|^2 -|\nabla \psi_g|^2$. For $g$
large, we deduce that $\rho_g$
 is an almost ground state of $F$.
 Among all the possible $\rho_0$'s which are ground states
 of $F$, the limit of $\psi_g$ when $g$
 is large should be such that $\psi_0=\sqrt{\rho_0}$  minimizes
  the kinetic energy $\int |\nabla \psi|^2$ among all
$\psi$ such that $\rho = |\psi|^2$ is a ground state of
 $F$. This implies that the support of $\psi_0$ is the union of
$n$ connected sets $A_i$, $1\leq i \leq n$, which satisfy
(\ref{promin}).

\begin{figure}[bt]
\begin{centering}
\includegraphics[width=65mm]{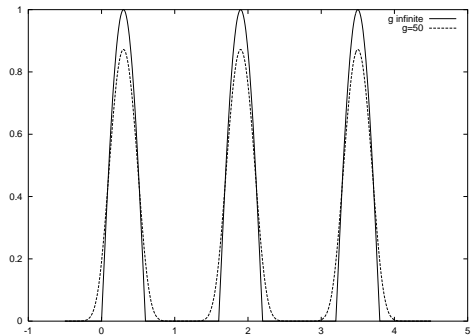}
\end{centering}
 \caption{Ground state of $E_g$ when   $g$ is large (dashed line), and its
  limiting profile $\psi_0$ (solid line). The bumps are of
  size $l$ and separated by a distance 1. }
 \label{fig:1}\vspace{-.2cm}
\end{figure}

\noindent {\bf More specific computations in 1D.} In dimension $1$
(for  $N$ atoms in a cylindrical annulus \cite{L}), that is if
$\calD=(0,L)$,
   then $n = n(\calD)=[L]+1$ (if $L$ is not an integer \cite{non-entier}),  and
  the $A_i$'s
   are intervals $(x_i,x_i+l)$, with $l=L/n-1+1/n$ and
  $x_i=i(l+1)$. Thus, $\psi_0(x)=\sqrt{(2L/nl)} \sin(\pi(x-x_i)/l)$ if $x\in
  (x_i,x_{i+1})$ and 0 otherwise (see figure~\ref{fig:1}).
 Moreover, $E_0(\psi_0) = \pi^2 L/ 2 l^2$. Indeed, the
ground state of $F$ provides $n$ sets $A_i$ separated from one
another by distance at least $1$. Hence $A_i\subset [a_i, b_i]$ for
all $i$, and $b_i +1 \le a_{i+1}$ for $i=1,\ldots , n-1$. Then,
since \eqref{Mstar} implies that $\int_{a_i}^{b_i} |u|^2  =
\frac{L}{n}$ for every $i$, $ \int |u'|^2 \ = \sum_{i=1}^{n}
\int_{a_i}^{b_i} |u'|^2 \ge \sum_{i=1}^{n} \frac {\pi^2}{(b_i -
a_i)^2}\int_{a_i}^{b_i} |u|^2  =
 \frac{L}{n}\sum_{i=1}^{n} \frac {\pi^2}{(b_i -
 a_i)^2},$
with equality if and only if the restriction of $u$ to each interval
$(a_i,b_i)$ is a scaled and normalized sine function multiplied by a
constant of modulus $1$. Moreover,  Jensen's inequality implies that
 $\sum \frac {1}{(b_i - a_i)^2} \ge \frac {n} {(n^{-1} \sum(b_i-a_i))^2 }
\ge   n/ l^2$, with equality if and only if $b_i-a_i = l$ for every
$i$.

One expects a boundary layer around each $A_i$. In this
one-dimensional setting, it is possible to compute it explicitly. In
order to do so, we assume that $\psi$ is a dilation of the limit
$\psi_0$, namely $\psi(x) = \sqrt\frac{2L}{(l+k)n} \sin
\left(\frac{\pi(x-i(l+1))}{l+k} \right)$ if $x\in [i(1+l) - k/2,
i(1+l)+l+k/2],$ and $\psi=0$ otherwise. The energy of this trial
function is computed explicitly, in the limit of small $k$:
$E_g(\psi) \approx \frac{\pi^2L}{2(l+k)^2} + \frac {gL^2} {4n} +
\frac g 4 A \left(\frac{k}{l+k}\right)^6,$ where $A = 13L^2
\pi^6/(90n).$ Minimizing this expression with respect to $k$ yields
$k = \left(({2L\pi^2l^3})/({3A}) \right)^{1/5} g^{-1/5}.$ Inserting
this into the expression of the energy, we find
$$E_g(\psi) \approx\frac{\pi^2L}{2l^2} + \frac {gL^2} {4n} - \frac56
\left(\frac{60L^4\pi^6n^2}{13 l^{12}} \right)^{1/5} g^{-1/5},$$ as
$g$ is large.

The above computation indicates that the boundary layer around each
bump of the limit function $\psi_0$ is of order $g^{-1/5}$ and if
$x$ denotes the scaled distance to the boundary, then the matching
 between $\psi_0$ and 0 in the boundary layer
  is described by the solution of $u''=cx^3u$. This boundary
layer decreases the energy by an amount of order $g^{-1/5}$.

\noindent {\bf Dimension 2 and 3.} There is no complete
determination of the $A_i$'s,
  except that once the sphere packing problem is known to provide a
   hexagonal lattice, the $A_i$'s are sets whose centers are located on
   an almost hexagonal lattice. Since minimizing $\lambda_1(A_i)$
   over $A_i$ with fixed volume implies that $A_i$ is a ball (see \cite{erdos}),
   condition (\ref{promin}) implies that $A_i$ "looks like" a ball
   to some extent. However, $\lambda_1(A_i)$ is increasing with
   respect to $A_i$, which implies that $A_i$ cannot be exactly a
   ball, but is closer to a hexagon (see figure~\ref{fig:2}).

A ground state $\psi_g$ of $E_g$ is close to $\psi_0$ in the $A_i$'s
and small in between. We need to understand better the smallness of
$\psi_g$. The Euler-Lagrange equation satisfied is $-\Delta \psi_g +
gW(x)\psi_g = 0,$ where $W(x) = U*|\psi_g|^2 - \lambda_g,$ and
$\lambda_g$ is the chemical potential. The shape of $U$ and the mass constraint on $\psi_g$
imply that $1\leq \max |W| \leq 2.$ Thus the Harnack inequality
\cite{GT} applied to the equation for $\psi_g$ yields
\begin{equation}\label{h1}
\inf \psi_g \geq 2gC_d e^{(-\sqrt{2g} T)} \max \psi_g,
\end{equation}
for some constant $C_d$ \cite{constante}, where $T$ is the size of
the period of $\rho_g$ and is of order $1$. This estimate is used
below in order to estimate the NCRIF.

 In the limit of very large $g$, the function $\psi_g$ is exponentially
 small: the potential $W$ is almost
equal to $W_0 = U*\rho_0 - \lambda_0$,  which vanishes in each
$A_i$, and is positive in between. Using appropriate
comparison solutions,  it is  possible to prove the estimate
$|\psi_g(x)| \leq \exp(-\delta \sqrt{g a_\delta}),$ for any $x$ such
that $\dist(x,A_i)>\delta$. The constant $a_\delta$ is the minimal
value of $W$ in this region, and is of order $ \delta^{(d+5)/2}$,
where $d$ is the dimension. The density is thus exponentially small
between the $A_i$'s. However, in the experiments, $g\sim 25$, so
that it is not  large to the point of having tiny density.

\noindent {\bf Small rotation} When the  sample is set under
rotation $\Omega$ about the $z$ axis, the free energy of the system
is defined by (\ref{eomega}) and $E_g$ is the energy defined in
(\ref{GP}). We assume that  the
  ground state $\psi$ of (\ref{eomega}) is of the form $\psi(x) =
  \sqrt{\rho_g(x)} e^{i\Omega S(x)}$ for small $\Omega$, where $\psi_g =
  \sqrt{\rho_g}$
   is a ground state of $E_g$, that is for $\Omega=0$. This equivalent
   to expanding the phase in terms of $\Omega$ and assume
   that the first order variation in the phase is not sensitive to the variations
   in density in terms of $\Omega$.  Then,
   the phase $S$ should minimize $\int \rho_g |\nabla S -{\bf
     e}_z\times {\bf r}|^2$ among all possible test functions. This provides an expansion
   of $e(\Omega)$ for small $\Omega$ and hence a value for $I$
   which allows to
  compute
  \begin{equation}\label{ncrif}
  \hbox{NCRIF}=\frac{\inf_{S}\int \rho_g |\nabla S-{\bf e}_z\times {\bf
  r}|^2}{\int \rho_g r^2}.\end{equation}
   Two limiting cases are easily identifiable:
  when $\rho_g=1$ (i.e when $g$ is small), this ratio is 1, and when
   $\rho_g$ is periodic and has all its mass
  localized in the center of the cell, this ratio tends to 0.
 For intermediate values of $g$,
  the wave function is localized
   in $A_i$,  with tails in between the sets which are small, but not too small.
   Then, (\ref{ncrif}) implies
\beq\label{lbn}\hbox{NCRIF}\geq \frac{\inf \rho_g}{\max
\rho_g}\frac{\inf_S\int |\nabla S - {\bf e}_z\times {\bf
    r}|^2 }{\int r^2} =\frac{\inf \rho_g}{\max
\rho_g}.\eeq The last equality is due to the fact that $\inf_S\int
|\nabla S - {\bf e}_z\times {\bf
    r}|^2$ is achieved for $\nabla S=0$.
Note that this estimate is a mere consequence of (\ref{ncrif}) and
not of the shape of $\rho_g$. The ratio $\max \rho_g/\inf \rho_g$
was estimated above \eqref{h1}. We thus have
\beq\label{lbn2}\hbox{NCRIF} \geq 4g^2C_d^2 e^{-2\sqrt{2g}T},\eeq
Since $\sqrt g$ is of order 5, this implies that $\hbox{NCRIF} \neq 0$ for
the experimental values.

   In the very large $g$ limit, (\ref{lbn2}) is not so good, but in
   this case,
     we may
  replace $\rho_g$ by $\rho_0$ in (\ref{ncrif}).
 Moreover, in each $A_i$,
   we can define local coordinates ${\bf r}_i$ with respect to a point
    in $A_i$ whose coordinate is ${\bf x}_i$. Then ${\bf r}={\bf r}_i+{\bf x}_i$
     and the phase $S$ can be defined as a local phase $S_i$ in each $A_i$
     through $\nabla S=\nabla S_i+{\bf x}_i^\perp$ where
     ${\bf x}_i^\perp={\bf e}_z\times {\bf x}_i$.
We thus have
$$\hbox{NCRIF} \approx\frac{ \sum_{i=1}^{n(\calD)} \inf_{S_i} \int_{A_i}
  \rho_0|\nabla S_i - {\bf e}_z\times {\bf
    r}_i|^2}{\int \rho_0 r^2}.$$
Assuming that each $A_i$ is the translation of a reference set $A_0$,
 the numerator is proportional to $n(\calD) $ times the infimum
  of the cell problem, which is always less
   than $V \ Vol(A_0)$.  Note that this cell problem
  depends on the volume since the size of $A_0$ depends
   on $n(\calD)$. If $V$ is large,
 a coarse-grained approximation for $\rho_0$ yields that
$\int \rho_0 r^2 \approx  \int r^2 \propto  V^{(d+2)/d}$, where $d$
is the dimension bigger than 2.
Hence $\hbox{NCRIF}\leq V^{-d/2}\ Vol(A_0)$, which tends to $0$ in the limit of
large $V$.
However, according to
Legget \cite{L}, the system can be considered as superfluid if
$\hbox{NCRIF} \gg 1/N$, where $N$ is the number of particles equal to $\nu
V)$ where $\nu$ is the initial average density (included in our
rescaling providing $g$). In a thermodynamic limit with $V$ large
$g$ large, and $\nu$ not fixed but large as well, we may still have
$\hbox{NCRIF} \gg 1/N.$

   Let us point out that this behaviour contrasts to  the 1D case,
   where, in the large $g$ asymptotic, the NCRIF is zero: indeed
   a similar computation yields that it is equal to $L^2/(\int \rho_g \int 1/\rho_g)$ (see also \cite{L}).
   In the large $g$ case, this tends to 0 since $\rho_g$ tends to $\rho_0$ which is
   compactly supported and thus $\int 1/\rho_0 = \infty$.

   \noindent
  {\bf Conclusion:} We have derived properties on the density
  of the ground state of a quantum crystal.  This has allowed us to estimate the NCRIF and
  find that the system is supersolid, on the basis of a definition of Leggett \cite{L}.
   This complements the results
   of \cite{PR} and provides theoretical justification of the
   non zero NCRIF in the experiments of Kim and Chan \cite{KC}.

\hfill

\noindent {\bf Appendix:} {\em Proof of Theorem 1. Step 1.} If
$\rho$ satisfies \eqref{Mstar}, we prove that  $F(\rho) =  {V^2}/{
n}$. Indeed, for every $j$, if $x\in A_j$, then $B_x \cap
(\cup_{k=1}^{ n} A_k) = A_j$, where $B_x$ is the ball of radius 1
centered at $x$.
 Since \eqref{Mstar} implies
that $\rho( \calD\setminus \cup A_j) = 0$,
 we get that
if $x\in A_j$, then $ \rho(B_x) = \rho(A_j) =  V/{ n}.$
 Since $\cup A_j$ is a set of full $\rho$ measure, we get that
  $F(\rho)=\int(U*\rho)({\bf r})\rho({\bf r})
 d{\bf r}=\sum \rho^2(A_j)=  {V^2}/{ n}$.

{\em Step 2}.  Let $\rho$ be a ground state for $F$.  We can argue
by induction that there exist $n=n({\cal D})$ points $x_1,\ldots,
x_n$ such that \beq \mbox{$|x_i - x_j| \ge 1$ and $U*\rho(x_i)
=\inf U*\rho=
   F(\rho) /V$}.
\label{goodpoints}\eeq The last equality is in fact the
Euler-Lagrange equation. The definition of $ n$ implies that if the
$x_i$ are any such points, then $\cup B_{ x_i} \supset\calD$. So
that $V\leq\rho( \cup_{i} B_{ x_i} ) \le \sum_{i} \rho(B_{ x_i} ) =
\sum_{i} U*\rho(x_i) = \frac{ n}{V}\ F(\rho).$ Thus $\min F=F(\rho)=
V^2/ n$.

{\em Step 3}. We have to check that (\ref{Mstar}) holds. For each
$x_j$,  we define $A_j = \{ x\in B_{x_j} \cap \mbox{supp}\,\rho\}$.
Then by the Euler-Lagrange equation, $U*\rho(x) = {V} / { n}$ in
$A_j$, hence $\rho(A_j) =V/n$.  We have that $ \rho(\cup A_i) = \sum
\rho(A_i)$, hence $\rho(A_i\cap A_j) = 0$ and if $y_i\in A_i$,  then
$U*\rho(y_i) = \rho(B_{y_i}) = V/ n$. Hence the points
$(\{x_1,\ldots, x_n\} \cup \{y_i\})\setminus \{ x_i\}$ satisfy
\eqref{goodpoints} and  this proves \eqref{Mstar}.

\hfill

\centerline{\bf Acknowledgments}

We are very grateful to  C.Josserand, Y.Pomeau, S.Rica  for
 their interest and remarks. We also wish to thank
 S.Balibar and A.L.Leggett for very fruitful discussions.


\begin{thebibliography}{99}
\bibitem{non-entier} If $L$
   is an integer,  one has to take
     $n=L$, and $\psi_0^2$ is a sum of Dirac masses at the points
   $x_i = i$. This example allows to
     see  that the size $l$ of the bumps gets  small
      as $L$ gets close to an integer, and decreases at least like $1/L$
       when $L$ gets large.
\bibitem{constante}  $C_d$  depends on the
  dimension:  $C_d = \omega_d / (2d^d)$, where
  $\omega_d$ is the volume of the $d$-dimensional unit ball. Thus $C_1=1$,
   $C_2 = \pi/8 $ and
  $C_3=2\pi/81.$
\end{thebibliography}
\end{document}